\documentclass{webofc}
\usepackage[varg]{txfonts} 

\def\beq{\begin{equation}}
\def\eeq{\end{equation}}
\def\beqa{\begin{eqnarray}}
\def\eeqa{\end{eqnarray}}

\begin{document}
\title{Single-top and top-antitop cross sections}

\author{\firstname{Nikolaos} \lastname{Kidonakis}\inst{1}\fnsep\thanks{\email{nkidonak@kennesaw.edu}} }

\institute{Department of Physics, Kennesaw State University, Kennesaw, GA 30144, USA}

\abstract{
I present high-order calculations, including soft-gluon corrections, for single-top and top-antitop production cross sections and differential distributions. For single-top production, results are presented for the three different channels in the Standard Model, for associated production with a charged Higgs, and for processes involving anomalous couplings. For top-antitop pair production, total cross sections and top-quark transverse-momentum and rapidity distributions are presented for various LHC energies.
}
\maketitle
\section{Introduction}

Higher-order soft-gluon corrections have been calculated through N$^3$LO for top-quark production via various processes; see Ref. \cite{NKtop} for a review. Here I present the latest results for $t$-channel and $s$-channel single-top production, $tW$ and $tH^-$ production, $tZ$ production via anomalous couplings, and $t{\bar t}$ production. For all these processes QCD corrections are very significant and are dominated by soft-gluon corrections.

I calculate and resum these soft corrections at next-to-next-to-leading logarithm (NNLL) accuracy for the double-differential cross section, which is then used to calculate top-quark transverse-momentum and rapidity distributions and total cross sections. Finite-order expansions at approximate NNLO (aNNLO) and approximate N$^3$LO (aN$^3$LO), matched to exact results, provide the best predictions for these quantities. 

We calculate soft-gluon corrections for partonic processes of the form 
$$
f_{1}(p_1)\, + \, f_{2}\, (p_2) \rightarrow t(p_t)\, + \, X 
$$
and we define
$s=(p_1+p_2)^2$, $t=(p_1-p_t)^2$, $u=(p_2-p_t)^2$,
and $s_4=s+t+u-\sum m^2$. At partonic threshold $s_4 \rightarrow 0$.
The $n$th-order soft-gluon corrections appear as $\left[\ln^k(s_4/m_t^2)/s_4\right]_+$ where $k \le 2n-1$.

Moments of the partonic cross section,
${\hat \sigma}(N)=\int (ds_4/s) \; e^{-N s_4/s} {\hat \sigma}(s_4)$, 
can be written in factorized form
$$
\sigma^{f_1 f_2\rightarrow tX}(N,\epsilon)
= H_{IL}^{f_1 f_2\rightarrow tX}\left(\alpha_s(\mu_R)\right) \, 
S_{LI}^{f_1 f_2 \rightarrow tX}\left(\frac{m_t}{N \mu_F},\alpha_s(\mu_R) \right) 
\prod  J_{\rm in} \left(N,\mu_F,\epsilon \right)
\prod J_{\rm out} \left(N,\mu_F,\epsilon \right)
$$ 
in $4-\epsilon$ dimensions, where $H_{IL}^{f_1 f_2\rightarrow tX}$ is a hard function and $S_{LI}^{f_1 f_2\rightarrow tX}$ is a soft-gluon function. $S_{LI}^{f_1 f_2 \rightarrow tX}$ satisfies the renormalization group equation
$$
\left(\mu \frac{\partial}{\partial \mu}
+\beta(g_s)\frac{\partial}{\partial g_s}\right)\,S_{LI}^{f_1 f_2 \rightarrow tX}
=-(\Gamma^\dagger_S)^{f_1 f_2 \rightarrow tX}_{LK} S_{KI}^{f_1 f_2 \rightarrow tX}-S_{LK}^{f_1 f_2 \rightarrow tX}(\Gamma_S)_{KI}^{f_1 f_2 \rightarrow tX}
$$
where the soft anomalous dimension $\Gamma_S^{f_1 f_2 \rightarrow tX}$ controls the evolution of the soft function, 
giving the exponentiation of logarithms of $N$. To achieve NNLL accuracy we need to calculate 
the relevant soft anomalous dimensions at two loops.

\section{Single-top production}

We now provide results for single-top production in the $t$-channel, $s$-channel, and via $tW$ production. Fixed-order results for these processes are known at NNLO for the $t$-channel \cite{NNLOtch,BGYZ,BGZ} and $s$-channel \cite{NNLOsch}, and at NLO for $tW$ production \cite{Zhu}. Here we provide results with soft-gluon corrections at aNNLO for $t$- and $s$-channel production \cite{NKsingletop}, and at aN$^3$LO for $tW$ production \cite{NKtW16}; these results are obtained from NNLL resummation \cite{NKsingletop,NKtW16,NKtW2l}. We use MMHT2014 NNLO pdf \cite{MMHT2014}.
\begin{figure}[h]
\centering
\includegraphics[width=58mm,clip]{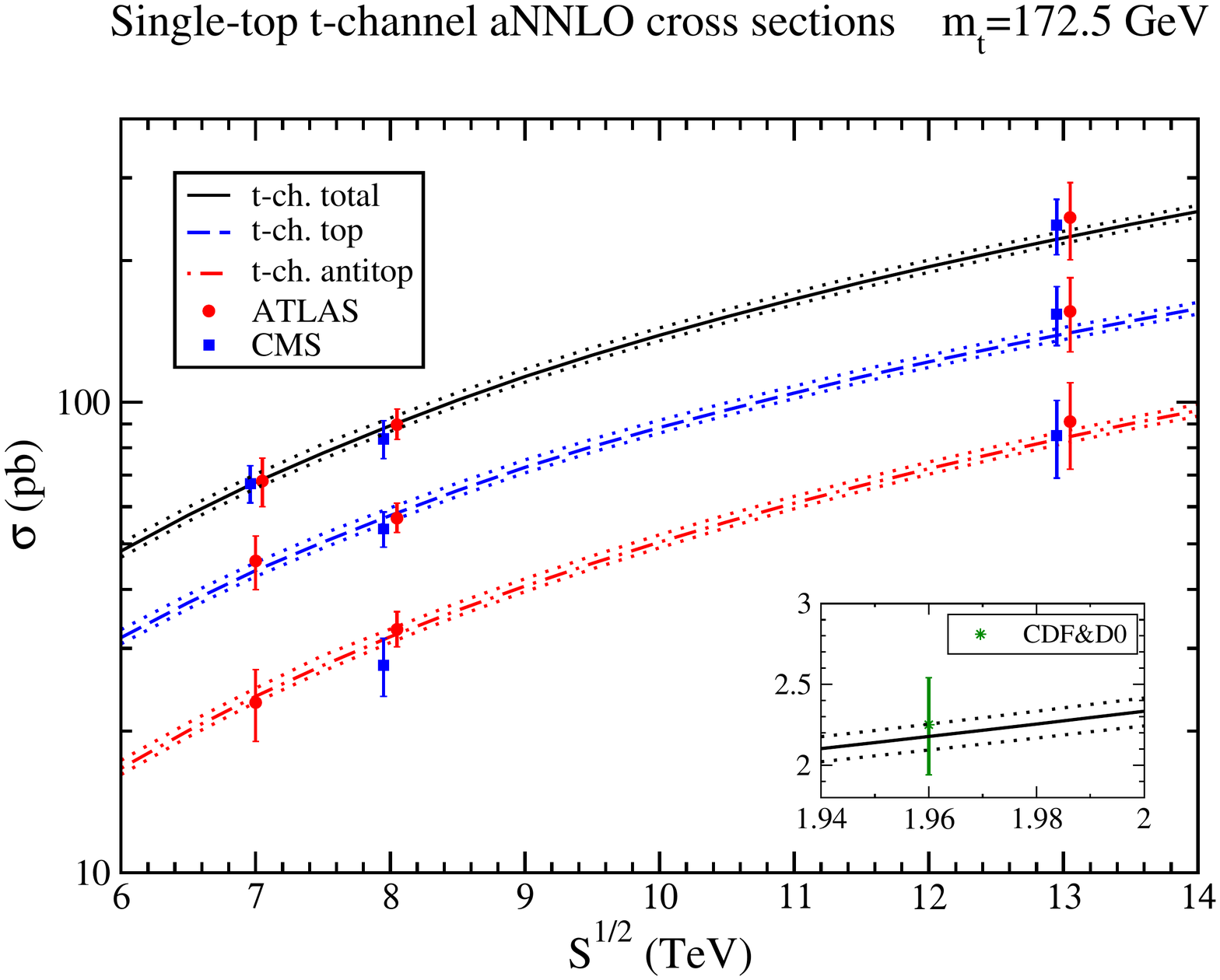}
\hspace{5mm}
\includegraphics[width=58mm,clip]{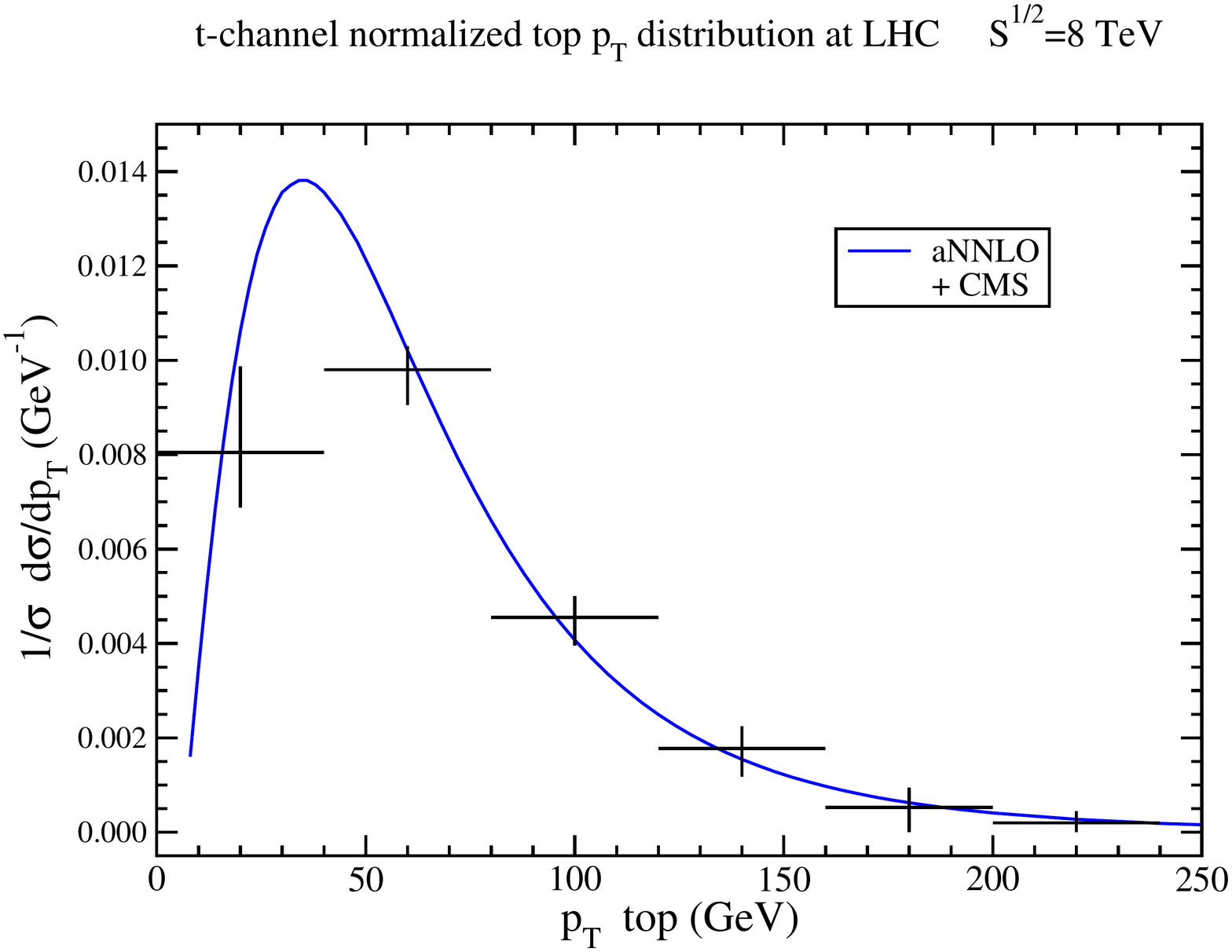}
\caption{(Left) Single-top $t$-channel aNNLO cross sections compared with CMS and ATLAS data at 7 TeV \cite{CMStch7,ATLAStch7}, 8 TeV \cite{CMStch8,ATLAStch8}, and 13 TeV \cite{ATLAStch13,CMStch13}, and with CDF and D0 combined data at 1.96 TeV \cite{CDFD0tch}. (Right) Top-quark aNNLO normalized $p_T$ distributions in $t$-channel production at 8 TeV compared to CMS \cite{CMStchpt8} data.}
\label{tch}
\end{figure}
\begin{figure}[h]
\centering
\includegraphics[width=58mm,clip]{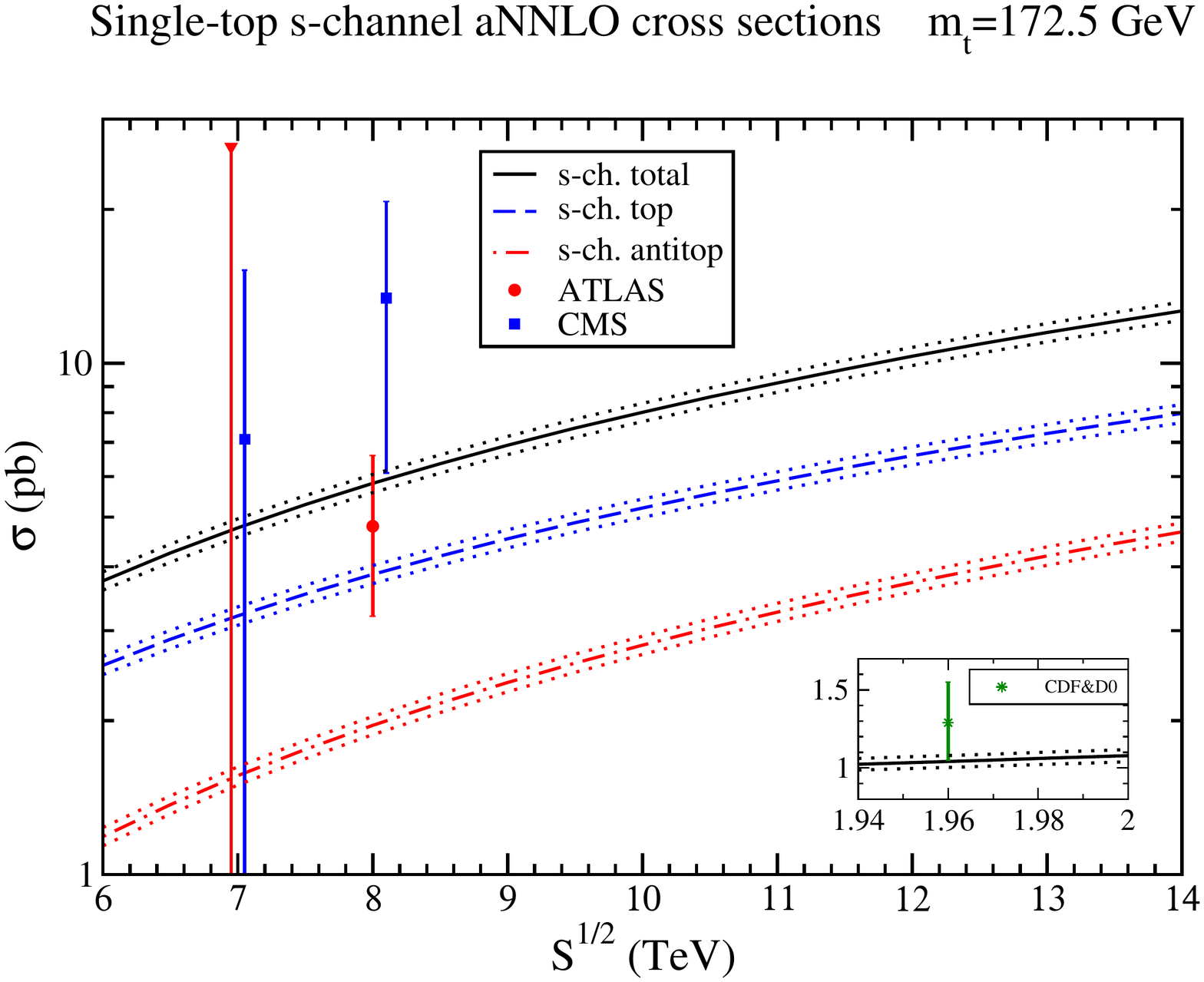}
\hspace{5mm}
\includegraphics[width=58mm,clip]{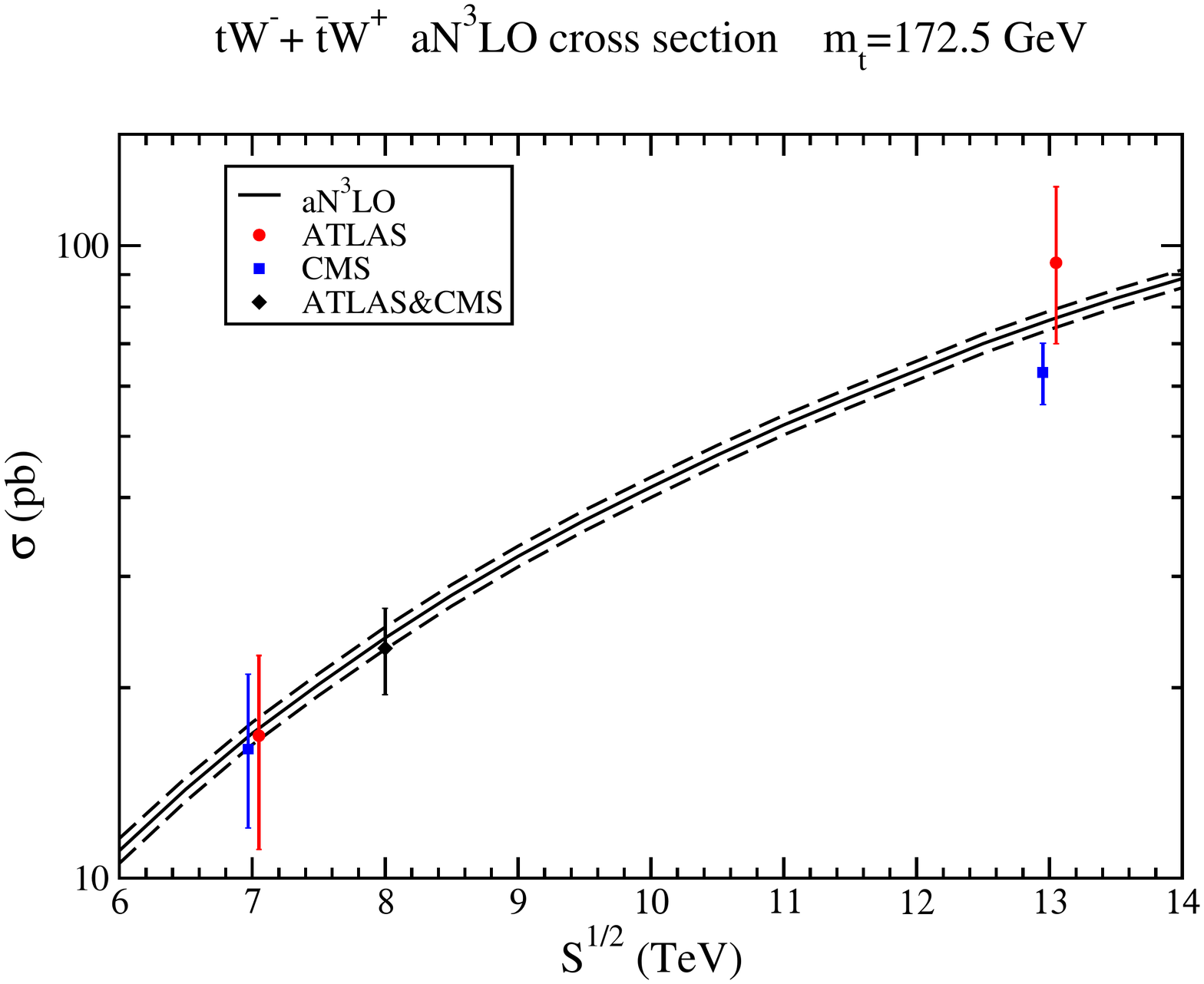}
\caption{(Left) Single-top $s$-channel aNNLO cross sections  
compared to ATLAS and CMS data at 7 TeV \cite{ATLASsch7,CMSsch7and8} and 
8 TeV \cite{CMSsch7and8,ATLASsch8}, and to CDF and D0 combined data 
\cite{CDFD0sch} at 1.96 TeV. (Right) aN$^3$LO cross sections for $tW$ production compared to ATLAS and CMS data at 7 TeV \cite{ATLAStW7,CMStW7}, 8 TeV \cite{ATLASCMStW8}, and 13 TeV \cite{ATLAStW13,CMStW13}.}
\label{stW}
\end{figure}

We begin with $t$-channel production at aNNLO. In the left plot of Fig. \ref{tch} we display total $t$-channel cross sections as functions of collider energy at the LHC and (inset) at the Tevatron. Results are given for the single-top cross section, the single-antitop cross section, and their sum. We observe very good agreement of the aNNLO theory curves \cite{NKsingletop,NKcipanp} with all available data from the LHC and the Tevatron at various energies.
The right plot of Fig. \ref{tch} shows the aNNLO normalized top-quark $p_T$ distributions at 8 TeV LHC energy, which describe the corresponding data from CMS \cite{CMStchpt8} quite well.

We continue with $s$-channel production at aNNLO. In the left plot of Fig. \ref{stW} we show cross sections for single-top and single-antitop $s$-channel production, and their sum, as functions of energy at the LHC and (inset) at the Tevatron. We again observe very good agreement of the aNNLO theory curves \cite{NKsingletop,NKcipanp} with available data at LHC and Tevatron energies.

We next discuss $tW$ production at aN$^3$LO. In the right plot of Fig. \ref{stW} we show the total $tW^-$+${\bar t}W^+$ cross section as a function of LHC energy. Very good agreement is observed between the aN$^3$LO results \cite{NKtW16,NKcipanp} and the data at 7, 8, and 13 TeV energies.

\section{$tH^-$ production}

We continue with $tH^-$ production in the MSSM or other two-Higgs-doublet models \cite{NKtH}. We use MMHT2014 NNLO pdf \cite{MMHT2014} for our numerical results.

\begin{figure}[h]
\centering
\includegraphics[width=58mm,clip]{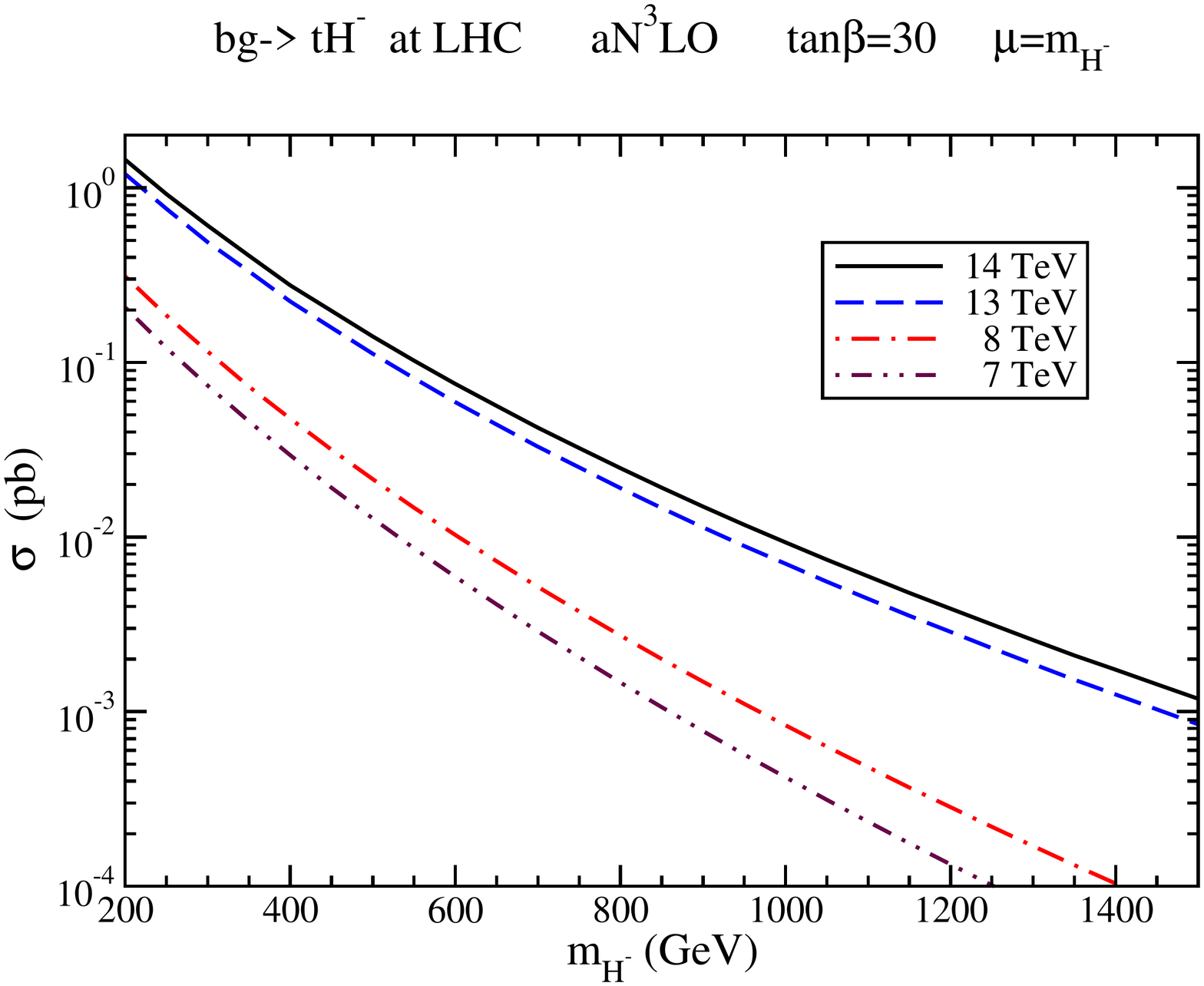}
\hspace{5mm}
\includegraphics[width=58mm,clip]{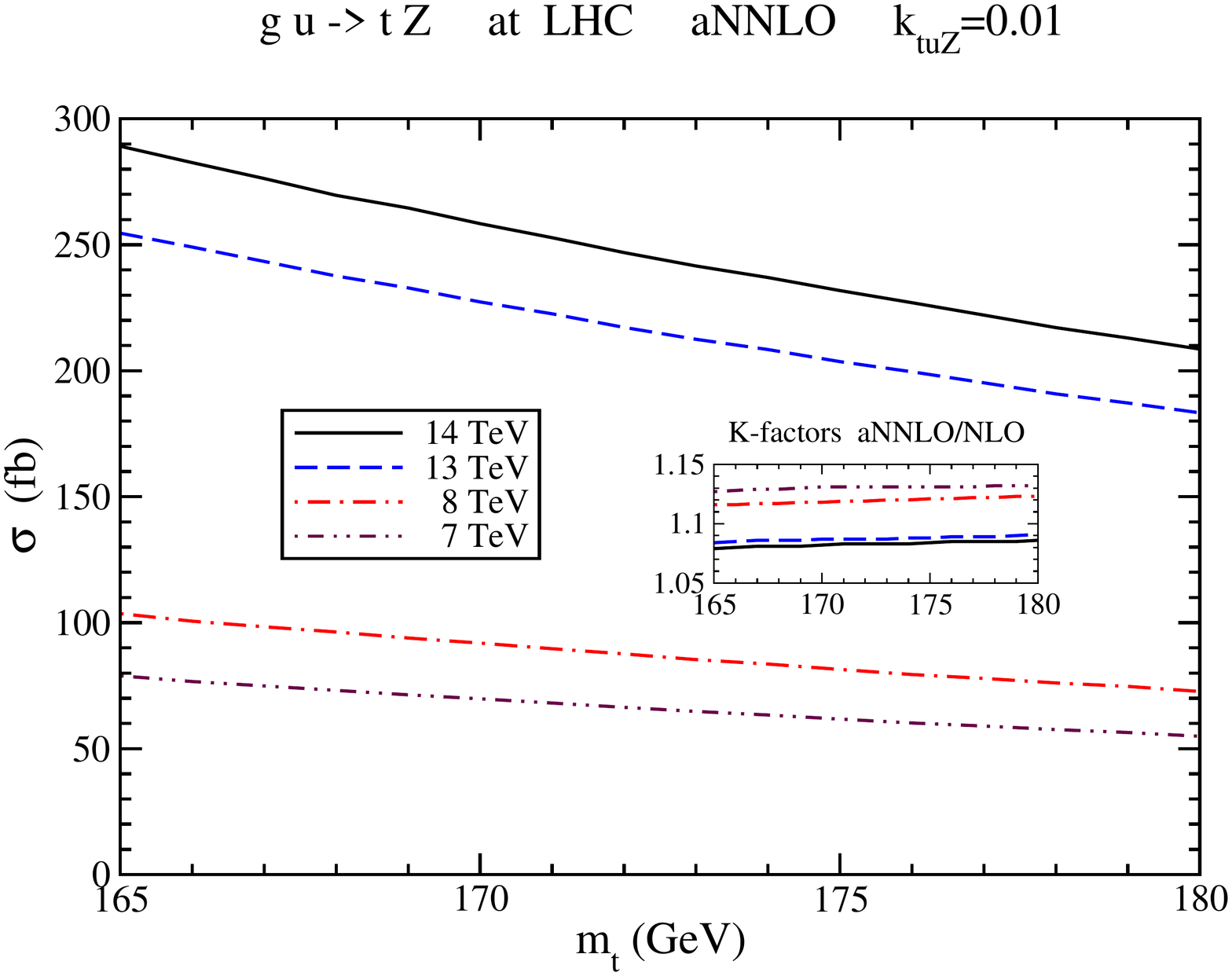}
\caption{(Left) aN$^3$LO cross sections for $tH^-$ production. 
(Right) aNNLO cross sections for $tZ$ production via anomalous couplings.}
\label{tHZ}
\end{figure}

In the left plot of Fig. \ref{tHZ} we show the aN$^3$LO total cross section for $tH^-$ production \cite{NKcipanp} as a function of charged-Higgs mass at LHC energies of 7, 8, 13, and 14 TeV. We use $\tan\beta=30$. The soft-gluon corrections are large for this process. Top-quark $p_T$ and rapidity distributions in this process have also been presented in \cite{NKtH}.

\section{$tZ$ production via anomalous couplings}

Next, we discuss soft-gluon corrections in $tZ$ production in models with anomalous $t$-$q$-$Z$ couplings \cite{fcnc,NKtZ}. The NLO corrections for this process were calculated in Ref. \cite{NLOtqZ}. The complete NLO corrections are very well approximated by the soft-gluon corrections at that order.

We use CT14 pdf \cite{CT14} for our numerical results in this process.
In the right plot of Fig. \ref{tHZ} we plot the aNNLO total cross section for $tZ$ production as a function of top-quark mass at LHC energies of 7, 8, 13, and 14 TeV. 

The $K$-factors shown in the inset plot show that the aNNLO corrections are large and significantly enhance the NLO cross section, especially at lower energies. This is important in providing theoretical input to experimental limits on the couplings \cite{CMStqZ,ATLAStqZ}. Top-quark differential distributions in this process have been presented in Ref. \cite{NKtZ}.

Similar results have more recently been presented for $t \gamma$ production via anomalous couplings in Ref. \cite{MFNK}.

\section{Top-antitop pair production}

Finally, we discuss top-antitop pair production \cite{NKtt,NKtt2}. The soft anomalous dimensions are $2 \times 2$ matrices for the $q{\bar q} \rightarrow t {\bar t}$ channel, and $3 \times 3$ matrices for the $gg \rightarrow t {\bar t}$ channel.

\begin{figure}[h]
\centering
\includegraphics[width=10cm,clip]{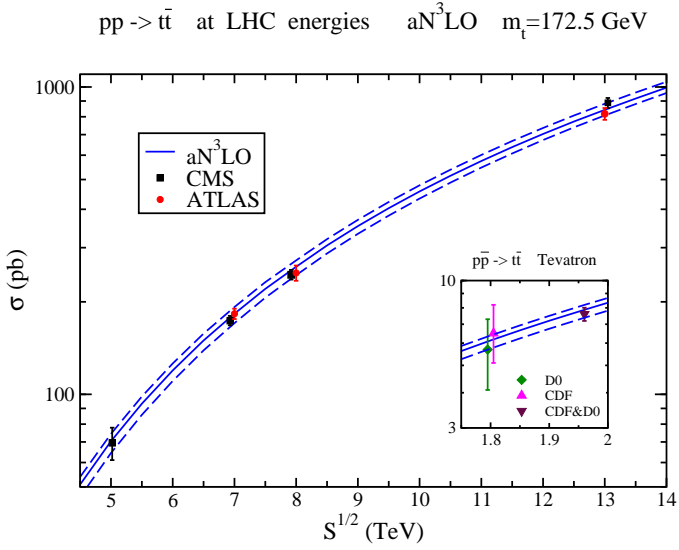}
\caption{Top-antitop aN$^3$LO cross sections compared with CMS data at 5.02 TeV \cite{CMS5.02} and with ATLAS and CMS data at 7 TeV \cite{ATLAStt7,CMStt7and8}, 8 TeV \cite{ATLAStt8,CMStt7and8}, and 13 TeV \cite{ATLAStt13,CMStt13} LHC energies. The inset shows the aN$^3$LO cross section compared with CDF \cite{CDF1.8} and D0 \cite{D01.8} data at 1.8 TeV, and CDF\&D0 combined data \cite{CDFD01.96} at 1.96 TeV Tevatron energy.}
\label{tt}
\end{figure}

At one loop for $q{\bar q} \rightarrow t{\bar t}$, the elements of the 
$\Gamma^{q{\bar q} \rightarrow t{\bar t}}_S$ matrix in an $s$-channel singlet-octet color basis are
\beqa
\Gamma^{q{\bar q} \rightarrow t{\bar t}\, (1)}_{11}&=&\Gamma_{\rm cusp}^{(1)} \, ,
\quad
\Gamma^{q{\bar q} \rightarrow t{\bar t}\, (1)}_{12}=
\frac{C_F}{C_A} \ln\left(\frac{t_1}{u_1}\right) \, , 
\quad 
\Gamma^{q{\bar q} \rightarrow t{\bar t}\, (1)}_{21}=
2\ln\left(\frac{t_1}{u_1}\right) \, ,
\nonumber \\ 
\Gamma^{q{\bar q} \rightarrow t{\bar t}\, (1)}_{22}&=&\left(1-\frac{C_A}{2C_F}\right)
\Gamma_{\rm cusp}^{(1)} 
+4C_F \ln\left(\frac{t_1}{u_1}\right)
-\frac{C_A}{2}\left[1+\ln\left(\frac{s m_t^2 t_1^2}{u_1^4}\right)\right] \, . 
\nonumber
\eeqa

At two loops for $q{\bar q} \rightarrow t{\bar t}$, the elements of the 
$\Gamma^{q{\bar q} \rightarrow t{\bar t}}_S$ matrix are
\beqa
\Gamma^{q{\bar q} \rightarrow t{\bar t}\,(2)}_{11}&=&\Gamma_{\rm cusp}^{(2)} \, ,
\quad 
\Gamma^{q{\bar q} \rightarrow t{\bar t}\,(2)}_{12}=
\left(\frac{K}{2}-\frac{C_A}{2} N_{2l}\right) \Gamma^{q{\bar q} \rightarrow t{\bar t} \,(1)}_{12} \, ,
\nonumber \\
\Gamma^{q{\bar q} \rightarrow t{\bar t} \,(2)}_{21}&=&
\left(\frac{K}{2}+\frac{C_A}{2} N_{2l}\right) \Gamma^{q{\bar q} \rightarrow t{\bar t} \,(1)}_{21} \, ,
\nonumber \\
\Gamma^{q{\bar q} \rightarrow t{\bar t} \,(2)}_{22}&=&
\frac{K}{2} \Gamma^{q{\bar q} \rightarrow t{\bar t} \,(1)}_{22}
+\left(1-\frac{C_A}{2C_F}\right)
\left(\Gamma_{\rm cusp}^{(2)}-\frac{K}{2}\Gamma_{\rm cusp}^{(1)}\right) \, .
\nonumber
\eeqa
See also Ref. \cite{NKtop} for more details. Here $\Gamma_{\rm cusp}$ is the cusp anomalous dimension \cite{NKloop}.

The soft-gluon corrections are large and they are excellent approximations to the complete corrections at both NLO and NNLO. The additional corrections at aN$^3$LO provide further significant enhancements and must be included for precision physics. Another approach for the aN$^3$LO corrections has appeared recently in Ref. \cite{JPCS}.

In our numerical results below we use the MMHT2014 NNLO pdf \cite{MMHT2014}. The total top-antitop cross sections at aN$^3$LO \cite{NKtt2} are shown in Fig. \ref{tt} and compared with data at Tevatron and LHC energies. We find remarkable agreement between theory and data at all energies.

\begin{figure}[h]
\centering
\includegraphics[width=58mm,clip]{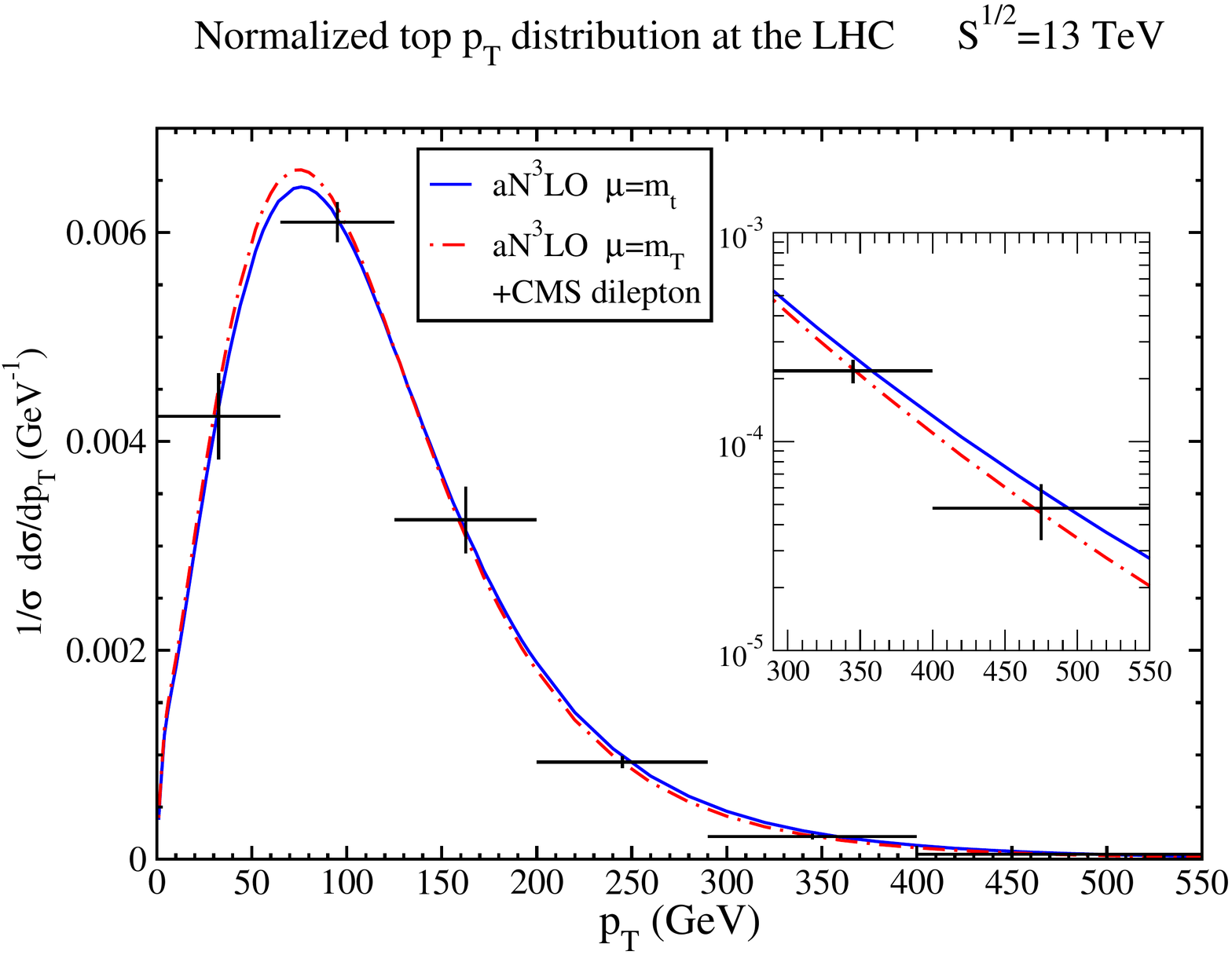}
\hspace{5mm}
\includegraphics[width=58mm,clip]{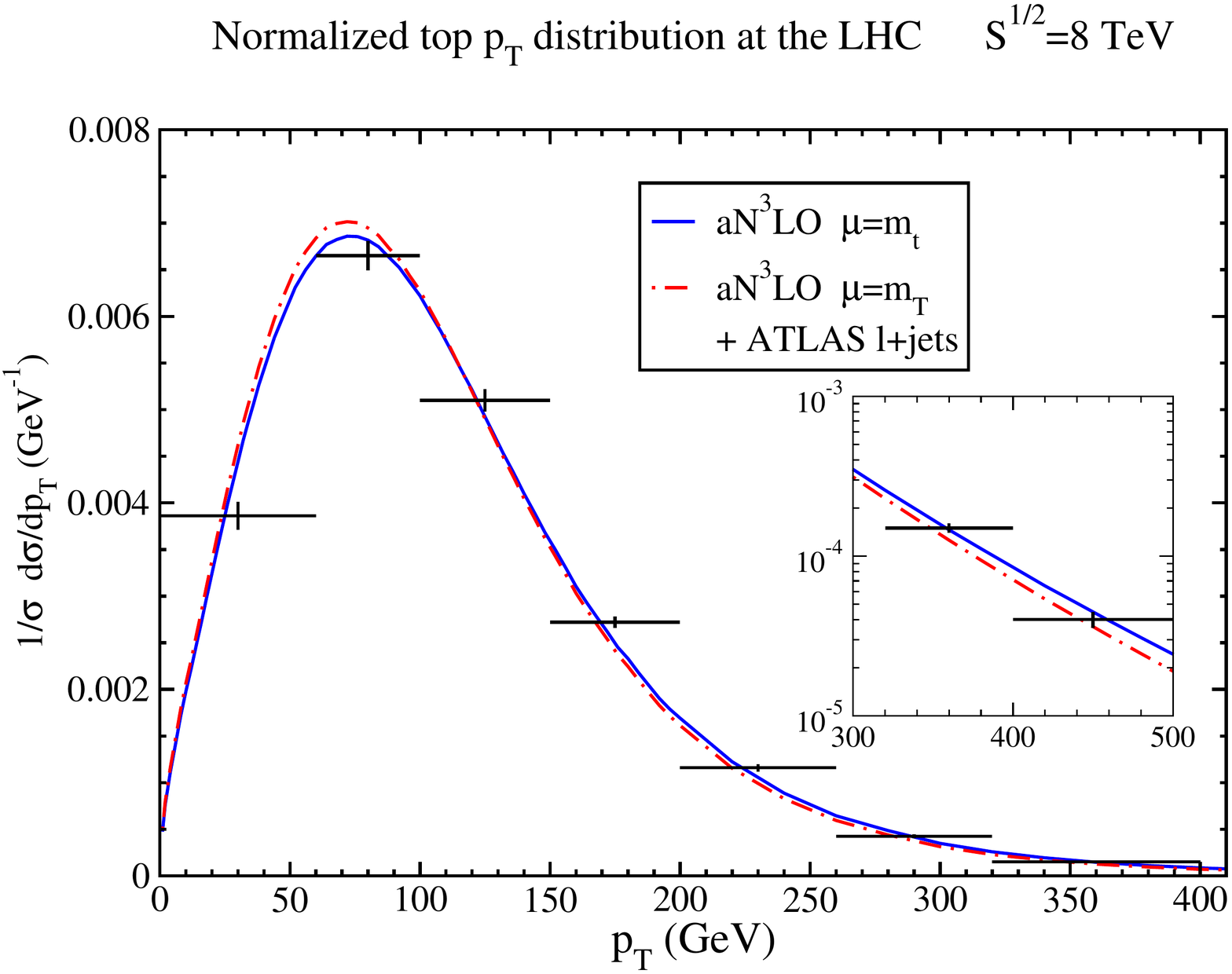}
\caption{aN$^3$LO top-quark normalized $p_T$ distributions (left) at 13 TeV energy compared with CMS \cite{CMSpty13} data and (right) at 8 TeV energy compared with ATLAS \cite{ATLASpty8} data.}
\label{ttpT}
\end{figure}

The aN$^3$LO top-quark normalized $p_T$ distributions in $t{\bar t}$ production are shown in Fig. \ref{ttpT} at 13 TeV (left plot) and 8 TeV (right plot) and compared with CMS and ATLAS data respectively. We find excellent agreement of the theoretical predictions with the data.

\begin{figure}[h]
\centering
\includegraphics[width=58mm,clip]{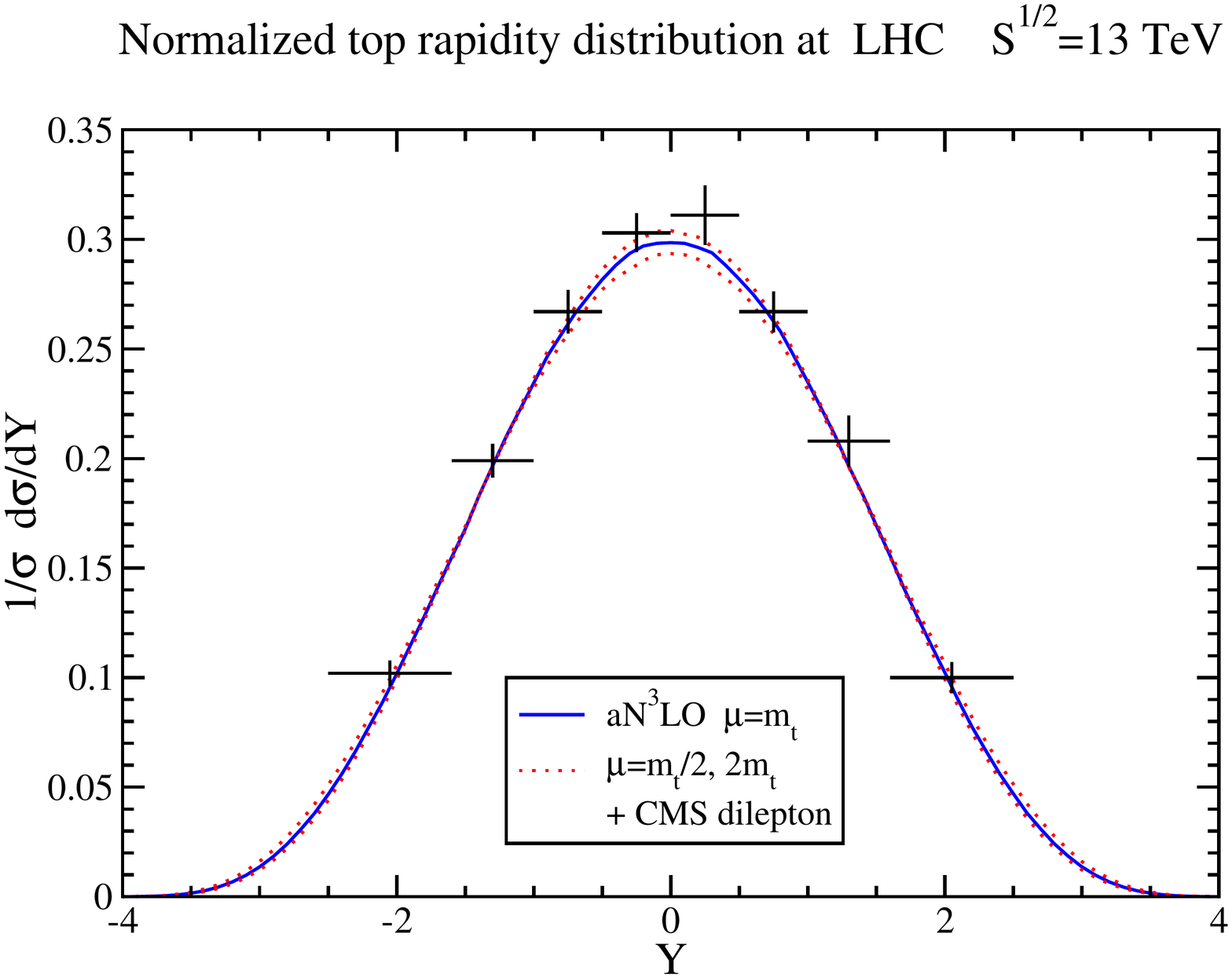}
\hspace{5mm}
\includegraphics[width=58mm,clip]{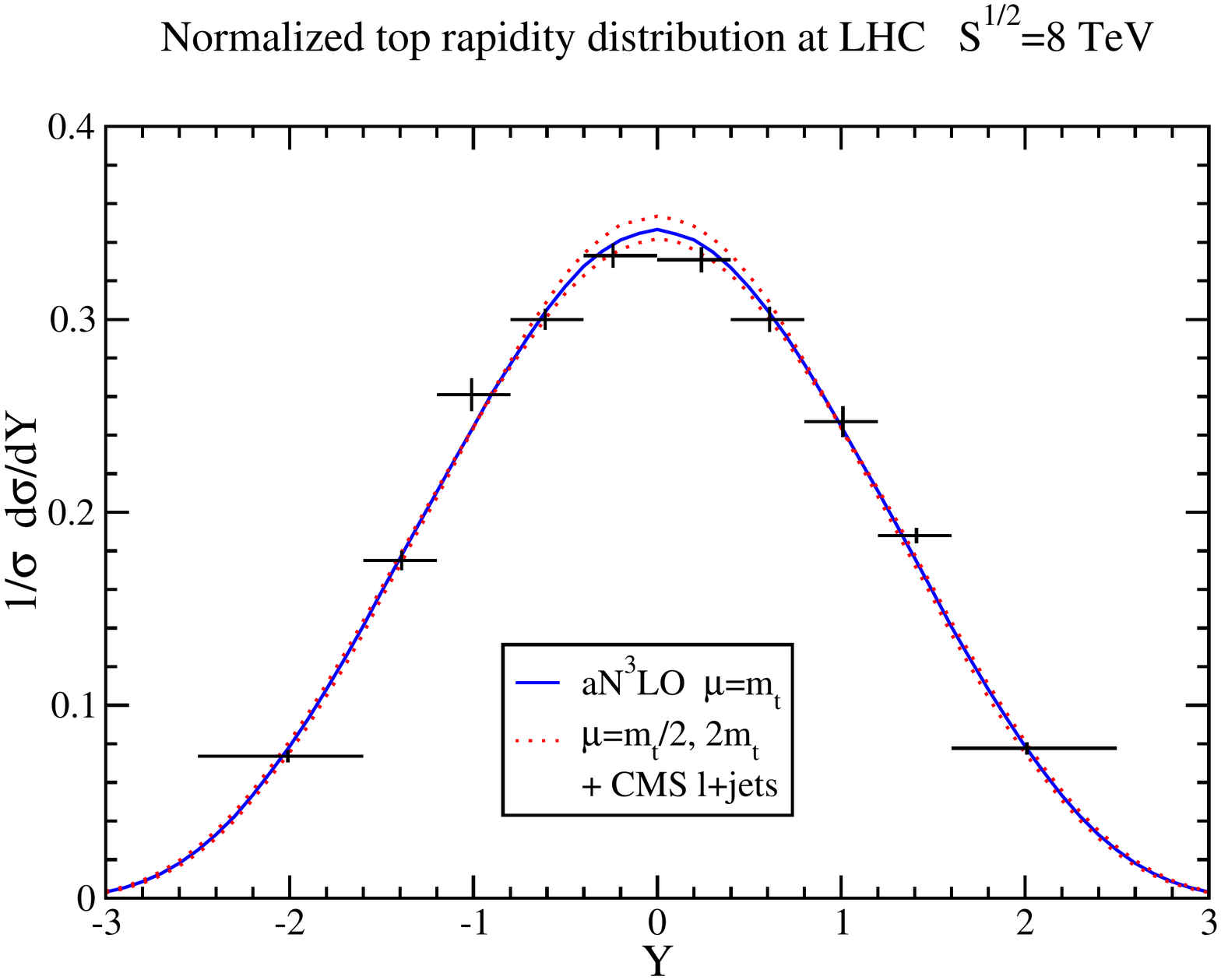}
\caption{Top-quark aN$^3$LO normalized rapidity distribution at (left) 13 TeV energy compared with CMS data \cite{CMSpty13} and (right) 8 TeV energy compared with CMS data \cite{CMSpty8}.}
\label{tty}
\end{figure}

The aN$^3$LO top-quark normalized rapidity distributions in $t{\bar t}$ production are shown in Fig. \ref{tty} at 13 TeV (left plot) and 8 TeV (right plot) and compared with CMS data. Again, we find that the theory curves provide an excellent description of the data.

\section{Summary}

We have discussed cross sections and distributions for various top-quark production processes. Soft-gluon corrections are important in all cases. We have shown results for $t$-channel and $s$-channel single-top production at aNNLO, $tW$ production at aN$^3$LO, $tH^-$ production at aN$^3$LO, $tZ$ production via anomalous couplings at aNNLO, and $t{\bar t}$ production at aN$^3$LO. We find excellent agreement with available collider data. The higher-order corrections are very significant and need to be included for better theoretical predictions. 

\section*{Acknowledgments}

This material is based upon work supported by the National Science Foundation under Grant No. PHY 1519606.

\end{document}